\documentclass{aa}  
\usepackage{graphicx}
\usepackage{txfonts}
\usepackage{hyperref}

\begin{document}

   \title{A circularly polarized low-frequency radio burst from the exoplanetary system HD~189733}

   % \subtitle{Exoplanets and UCDs with NenuFAR}

   \author{X. Zhang\inst{1}
            \and
            P. Zarka\inst{1,2}
            \and 
            J. N. Girard\inst{1}
            \and
            C. Tasse\inst{3}
            \and
            A. Loh\inst{1}
            \and
            E. Mauduit\inst{1,2}
            \and
            F. G. Mertens\inst{3,4}
            \and 
            E. Bonnassieux\inst{1,5,6}
            \and
            C. K. Louis\inst{1}
            \and
            J-M. Grießmeier\inst{2,7}
            \and
            J. D. Turner\inst{8,9}
            \and
            L. Lamy\inst{1,2,10}
            \and
            A. Strugarek\inst{11}
            \and
            S. Corbel\inst{11}
            \and
            B. Cecconi\inst{1,2}
            \and
            O. Konovalenko\inst{12}
            \and
            V. Zakharenko\inst{12}
            \and
            O. Ulyanov\inst{12}
            \and 
            P. Tokarsky\inst{12}
            \and
            M. Tagger\inst{7}
          }

   \institute{LIRA, Observatoire de Paris, Université PSL, Sorbonne Université, Université Paris Cité, CY Cergy Paris Université, CNRS, 92190 Meudon, France\\
        \email{Xiang.Zhang@obspm.fr}
        \and
        Observatoire Radioastronomique de Nançay (ORN), Observatoire de Paris, Université PSL, Univ Orléans, CNRS, 18330 Nançay, France
        \and
        LUX, Observatoire de Paris, Université PSL, Sorbonne Université, CNRS, 75014 Paris, France
        \and
        Kapteyn Astronomical Institute, University of Groningen, PO Box 800, 9700 AV Groningen, The Netherlands
        \and
        Instituto de Astrofísica de Andalucía, Consejo Superior de Investigaciones Científicas (CSIC), Glorieta de la Astronomía s/n, E-18008, Granada, Spain
        \and
        INAF, via P. Gobetti 101, I-40129 Bologna, Italy
        \and
        LPC2E, OSUC, Univ Orleans, CNRS, CNES, Observatoire de Paris, F-45071 Orleans, France
        \and
        Department of Astronomy and Carl Sagan Institute, Cornell University, Ithaca, NY, USA
        \and
        NHFP Sagan Fellow, USA
        \and
        Aix Marseille Université, CNRS, CNES, LAM, Marseille, France
        \and 
        Université Paris Cité and Université Paris Saclay, CEA, CNRS, AIM, 91190 Gif-sur-Yvette, France 
        \and
        Institute of Radio Astronomy, Mystetstv St. 4, 61002, Kharkiv, Ukraine
        }

   \date{Received September 15, 1996; accepted March 16, 1997}

% \abstract{}{}{}{}{} 
% 5 {} token are mandatory
 
  \abstract
   {}
   {We aim to detect low-frequency radio emission from exoplanetary systems, which can provide insights into planetary magnetic fields, star-planet interactions, stellar activity, and exo-space weather. The HD~189733 system, hosting a well-studied hot Jupiter, is a prime target for such searches.}
   {We conducted NenuFAR imaging observations in the 15–62~MHz range, in order to cover the entire orbital phase of HD~189733~b. Dynamic spectra were generated for the target and other sources in the field, followed by a transient search in the time-frequency plane. The data processing pipeline incorporated direction-dependent calibration and noise characterization to improve sensitivity. We also searched for periodic signals using Lomb-Scargle analysis.}
   {A highly circularly polarized radio burst was detected at 50~MHz with a flux density of 1.5 Jy and a significance of 6$\sigma$ at the position of HD~189733. No counterpart was found in Stokes I, likely because the emission is embedded in confusion noise and remains below the detection threshold. The estimated minimum fractional circular polarization of 38\% suggests a coherent emission process. A periodicity search revealed no weaker signals linked to the planet’s orbital period, the star’s rotational period, or the synodic period and harmonic period between them. The burst’s properties are consistent with cyclotron maser instability (CMI) emission, but the origin is still ambiguous. The comparison with theoretical models suggests star-planet interaction or stellar activity as potential origins. However, alternative explanations such as contamination from other sources along the line of sight (e.g. the companion M dwarf) or noise fluctuation cannot be ruled out. }
   {}


    \titlerunning{Circular burst from HD~189733}
    \authorrunning{X. Zhang et al}

    \maketitle
%
%-------------------------------------------------------------------

\section{Introduction}

The study of magnetic fields in exoplanetary systems is a frontier in astrophysics, offering profound insights into multiple planetary characteristics and phenomena \citep{2015aska.confE.120Z, 2016kiss.rept....1L, 2019BAAS...51c.135L}. Magnetic fields provide key windows into the interior structures and dynamos of exoplanets, in analogy to Jupiter and other solar system planets \citep{2011A&A...531A..29H, 2021JGRE..12606724B}. Furthermore, understanding the magnetospheres of exoplanets is essential for grasping the complex dynamics of star-planet interactions (SPI), which can significantly influence planetary environments \citep{2004ApJ...602L..53I, 2007P&SS...55..598Z, 2009A&A...505..339L, 2010ApJ...722.1547S, 2015ApJ...815..111S, 2024NatAs...8.1359C, Zarka2025}. The presence and nature of magnetic fields are also pivotal in assessing the habitability of exoplanets, as they can shield atmospheres from stellar wind and cosmic radiation, and possibly mitigate atmospheric escape \citep{2010IJAsB...9..273H, 2014A&A...570A..99S, 2020JGRA..12527639G, 2021MNRAS.508.6001C}. Lastly, the detection and study of exoplanetary magnetic fields represent a potential avenue for the discovery and characterization of exoplanets, offering methods to infer the presence of planets beyond traditional transit and radial velocity techniques \citep{2018haex.bookE...9L}. 

Magnetic fields within exoplanetary systems can be probed through various observational methods, among which radio observations stand out as particularly advantageous. This method enables the direct constraint of magnetic field amplitudes without relying on complex modelling assumptions and exhibits a reduced susceptibility to false positives \citep{2015ASSL..411..213G, Zarka2025}. The predominant mechanisms for planetary radio emissions are Cyclotron Maser Instability (CMI) \citep{1998JGR...10320159Z, 2006A&ARv..13..229T} and synchrotron radiation \citep{2004P&SS...52.1449D}, with a current research emphasis on CMI due to its high efficiency. 

CMI emissions are highly circularly polarized, strongly beamed, and temporally variable \citep{2004JGRA..109.9S15Z}. They are produced near the local electron cyclotron frequency (\(f_{\rm ce}\)), which directly links the observed emission to the amplitude of the local magnetic field strength at the source of the emission\citep{2015aska.confE.120Z}. The emission spectrum shows a sharp cutoff at a maximum frequency set by the strongest magnetic field, typically near the planetary surface. For CMI to operate, the local plasma frequency \(f_{\rm pe}\) must satisfy \(f_{\rm pe}/f_{\rm ce} < 0.1\)\citep{2006A&ARv..13..229T}; otherwise, the emission is suppressed. Furthermore, CMI is strongly anisotropic and beamed along a narrow hollow cone at large angles (70°-90°) from the local magnetic field direction \citep{1986JGR....9113569P, 2008GeoRL..3513107H, 2017GeoRL..44.9225L}. This beaming geometry implies that detection depends on the observer's viewing angle.

The quest to detect radio emissions from exoplanets has spanned several decades, predominantly yielding non-detections \citep{2000ApJ...545.1058B,2004P&SS...52.1469F,2013ApJ...762...34H,2015MNRAS.446.2560M,2017MNRAS.467.3447L,2018MNRAS.478.1763L,2018A&A...612A..52O,2020A&A...644A.157D}. Despite these challenges, there have been multiple reports of tentative detections, such as 150~MHz radio emissions possibly emanating from exoplanets observed with the Giant Metrewave Radio Telescope (GMRT) \citep{2013A&A...552A..65L, 2014A&A...562A.108S}, alongside GHz-band radio pulses that may have planetary origins \citep{2018ApJ...857..133B}. Recent years have seen the potential identification of auroral emissions, reminiscent of star-planet interactions, utilizing the Low-Frequency Array (LOFAR) and the Australia Telescope Compact Array (ATCA) \citep{2020NatAs...4..577V,2021NatAs...5.1233C,2021A&A...645A..77P,2021A&A...645A..59T}. 
In particular, \citet{Tasse2025} have developped a new approach to generate massive amount of dynamic spectra from interformetric data. For the first time, \citet{Tasse2025} detected two bursts originated from stars hosting exoplanets, potentially revealing the first radio signatures of star-planet interactions via CMI.
These intriguing observations underscore the need for meticulous follow-up studies to ascertain the planetary nature of the detected signals.

HD~189733, a binary star system located 19 parsecs from Earth, has emerged as a compelling target for radio astronomical studies, particularly due to its strong magnetic field and prior tentative detections. The system comprises a K-type primary star (HD~189733~A) and an M dwarf companion (HD~189733~B), separated by approximately 216 AU \citep{2006ApJ...641L..57B}. Orbiting the primary is a hot Jupiter, HD~189733~b, characterized by a short orbital period of 2.219 days and a semi-major axis of $0.0313 \pm 0.0004$ AU, with mass and radius estimates of $1.15 \pm 0.04$~M$_J$ and $1.26 \pm 0.03$~R$_J$\footnote{M$_J$ and R$_J$ denote the mass and radius of Jupiter, respectively.}, respectively \citep{2005A&A...444L..15B}.
The orbital inclination of the planet with respect to the observer’s line of sight is approximately $85.7^\circ$ \citep{2009A&A...495..959B}, indicating an almost edge-on orbit. Spectroscopic measurements have shown that the planet’s orbital plane is closely aligned with the stellar equator, with a projected spin–orbit angle of only $1.4^\circ$ \citep{2016ApJ...817..106B}. This implies that the planet transits near the stellar equator, and that our line of sight is nearly in the equatorial plane of the star.

Observational efforts have focused on delineating the magnetic properties of both the star and the orbiting exoplanet. Zeeman–Doppler imaging has mapped the stellar surface magnetic field, revealing strengths up to 40 G, and further monitoring has indicated variations in the planetary magnetic field correlating with stellar activity \citep{2010MNRAS.406..409F, 2017MNRAS.471.1246F}. Archival analyses of Ca II K spectral lines have detected modulations consistent with the exoplanet’s orbital period, suggesting star-planet interactions (SPI) \citep{2018AJ....156..262C}. This hypothesis is supported by X-ray observations from XMM-NEWTON and Chandra, highlighting the primary star's high activity level and discrepancies in magnetic activity that may result from tidal interactions between the primary star and the exoplanet HD~189733~b \citep{2010ApJ...722.1216P, 2013ApJ...773...62P, 2014ApJ...785..145P, 2020MNRAS.493..559B, 2022A&A...660A..75P}. % Besides magnetic properties, astronomers also detected water vapour within the atmosphere of the exoplanet \citep{2021AJ....162..233B}.

In the radio domain, a 2.7 $\sigma$ detection at 150~MHz by GMRT has been reported, though the sensitivity was insufficient to measure time variations allowing to associate the emission with the planet \citep{2011A&A...533A..50L}. Stellar wind modelling, informed by the measured magnetic field maps, suggests that planetary emissions should occur at 2-25~MHz; however, emissions below 21~MHz may be obstructed by the stellar wind \citep{2019MNRAS.485.4529K}. Additional stellar wind modelling suggests that a stretch-and-break SPI mechanism \citep{2013A&A...557A..31L} could explain the observed signals \citep{2022MNRAS.512.4556S}, although further work on the applicability of the model \citep{2024NatAs...8.1359C} and observations covering the entire orbital phase of the exoplanet are needed to confirm these findings. However, subsequent multi-wavelength analysis and observations, including some with Arecibo, caution that these tentative detections might be false positives, attributing the observed magnetic enhancements to stellar activity rather than to SPI \citep{2019ApJ...872...79R, 2019ApJ...887..229R}.

To delve deeper into the potential SPI scenario, we conducted NenuFAR observations covering the full orbital phase of HD~189733~b across a broad low-frequency range. This approach is well suited to detecting beamed, low-frequency CMI emission induced by SPI, while also covering potential stellar plasma emission. Section \ref{observation} outlines the observational parameters and our comprehensive approach to covering the entire orbital phase of the system. Section \ref{process} details the data processing pipeline used to calibrate, image, and analyze the dataset. Our results are presented in Section \ref{sec:results}, including the detection of a highly circularly polarized burst and a search for periodic signals. In Section \ref{sec:discussion}, we discuss possible origins of the detected burst, considering star-planet interaction, intrinsic stellar activity, and contamination from background sources. Finally, Section \ref{sec:conclusions} summarizes our findings and outlines recommendations for future observations.

%--------------------------------------------------------------------
\section{Observations}
\label{observation}
Our study utilized the NenuFAR array in Nançay \citep[New extension in Nançay upgrading LOFAR; see][]{2012sf2a.conf..687Z, zarka2015nenufar, zarka:hal-04056720}. Designed as a low-frequency phased array and interferometer, NenuFAR operates within the 10-85~MHz range. The array is organized into hexagonal clusters of 19 dual-polarization antennas, each cluster termed a Mini-Array (MA). The MAs are analog-phased, which results in a steerable instantaneous Field of View (FoV) of 8° (85 MHz) to 60° (10 MHz). As of mid-2023 (the epoch of our observations), the array configuration includes 80 MAs positioned within a core area spanning 400 meters, complemented by 4 remote MAs situated at distances reaching up to 3 kilometers.

Between June and October 2023, we conducted an observational campaign targeting the HD~189733 system using NenuFAR. These observations were conducted in a dual-mode setup, enabling simultaneous imaging and beamformed data collection. However, this paper focuses solely on the analysis of imaging data\footnote{In comparison, beamformed data provide higher time/frequency resolution, but the noise level is also higher.}. The specific parameters for the imaging observations are detailed in Table \ref{tab:obs}. To minimize the effect of ionospheric disturbance at the lower end of the frequency spectrum, observations were conducted at night, when the target's elevation exceeded 40 degrees. Each night's observation spanned several hours, adopting a tracking mode to maintain the target at the pointing centre. For each observation, a 15-minute slot was allocated to observing a calibrator, either Cyg A or Cas A, immediately before or after the primary observation period. 

\begin{table}
\caption{Observational Parameters for the NenuFAR Campaign Targeting HD~189733.}
\label{tab:obs}
\centering
\begin{tabular}{c c c}
\hline\hline
Parameter & Value & Units \\
\hline
Array Configuration & Core + Remote MAs & \\
Frequency Range & 15.8 to 62.5 & MHz \\
Frequency Resolution \tablefootmark{a} & 15 & kHz \\
Temporal Resolution & 8 & seconds \\
Field of View (FoV) \tablefootmark{b} & 20 & degrees \\
Spatial Resolution \tablefootmark{c}& 0.5 & degrees \\
\hline
\end{tabular}
\tablefoot{
    \tablefoottext{a}{The frequency and temporal resolutions listed here correspond to the Level 1 data products, obtained after initial preprocessing steps that involve averaging, and thus differ from the raw data resolution.}
    \tablefoottext{b}{The FoV is calculated as the full width at half maximum (FWHM) of the angular resolution corresponding to a single MA of NenuFAR at 39~MHz.}
    \tablefoottext{c}{The spatial resolution is derived at 39~MHz. It is larger than the theoretical resolution of NenuFAR (0.1 degree) at the frequency, due to two of the four remote MAs being ﬂagged during most of the observational campaign.}
    }
\end{table}

\begin{figure}
\centering
\includegraphics[width=\hsize]{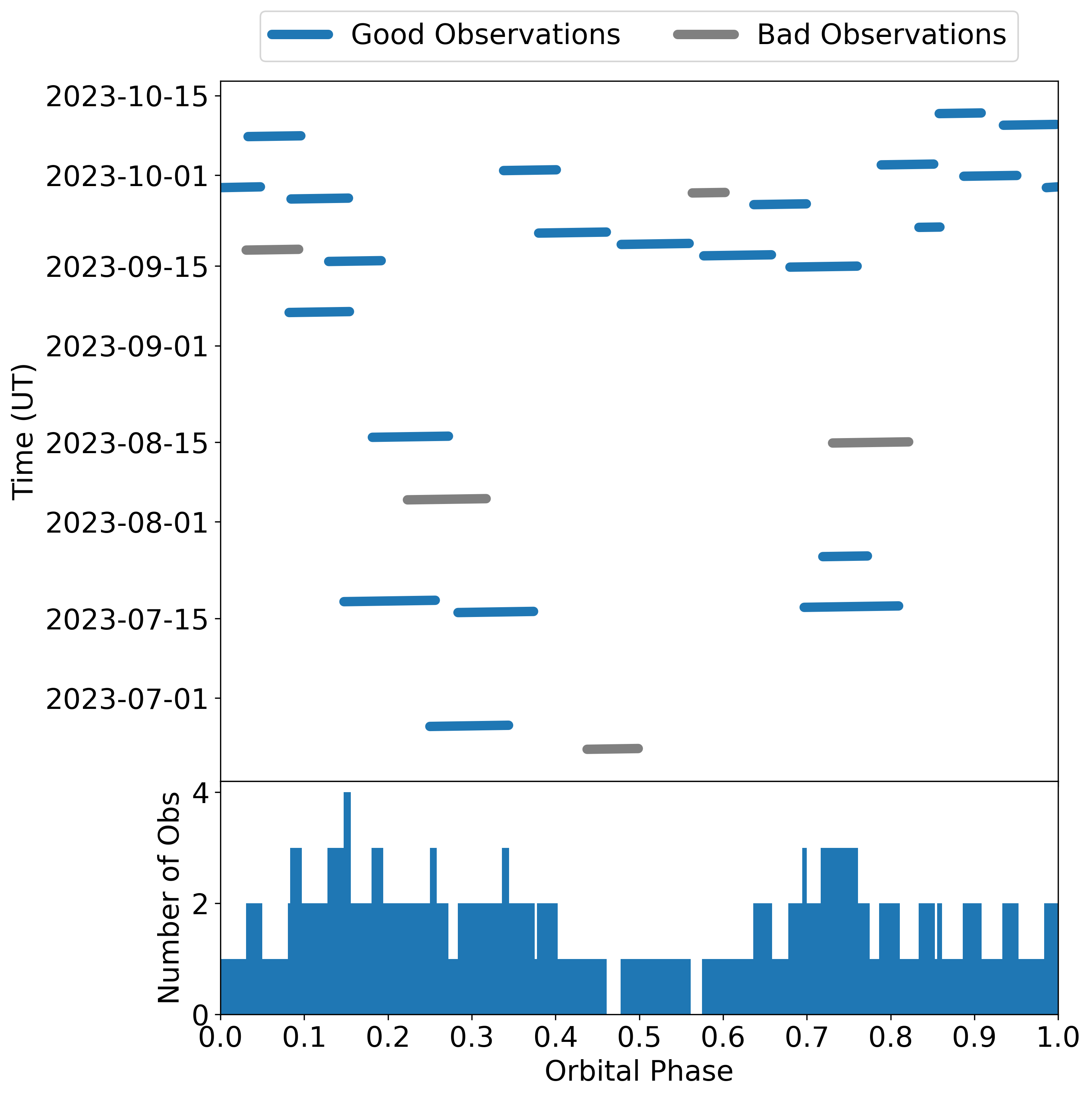}
\caption{Orbital Phase Coverage of HD~189733~b Observations. HD~189733~b is a transiting exoplanet, with the orbital phase of 0 defined as the moment of its transit \citep{2022ApJS..259...62I}. Observations marked in gray represent the data excluded from scientific analysis due to poor quality.}
\label{fig:phase}
\end{figure}

To comprehensively investigate the emissions from HD~189733~b, our observational campaign was carefully planned to span the exoplanet's full orbital phase. This strategy stems from the anticipation that CMI emissions are beamed, becoming detectable from Earth only during specific segments of the orbital cycle\citep{2023pre9.conf03091L}. The exoplanet's orbital parameters were derived from transit timing observations by the TESS mission \citep{2022ApJS..259...62I}. Figure \ref{fig:phase} illustrates the orbital coverage achieved by our campaign. Over a cumulative observation duration of 103 hours, our observations covered most of HD~189733~b’s orbital phase, despite minor gaps introduced by the removal of low-quality data (see Section \ref{sec:quality}).

\section{Data reduction}
\label{process}

\begin{figure*}
\centering
\includegraphics[width=\textwidth]{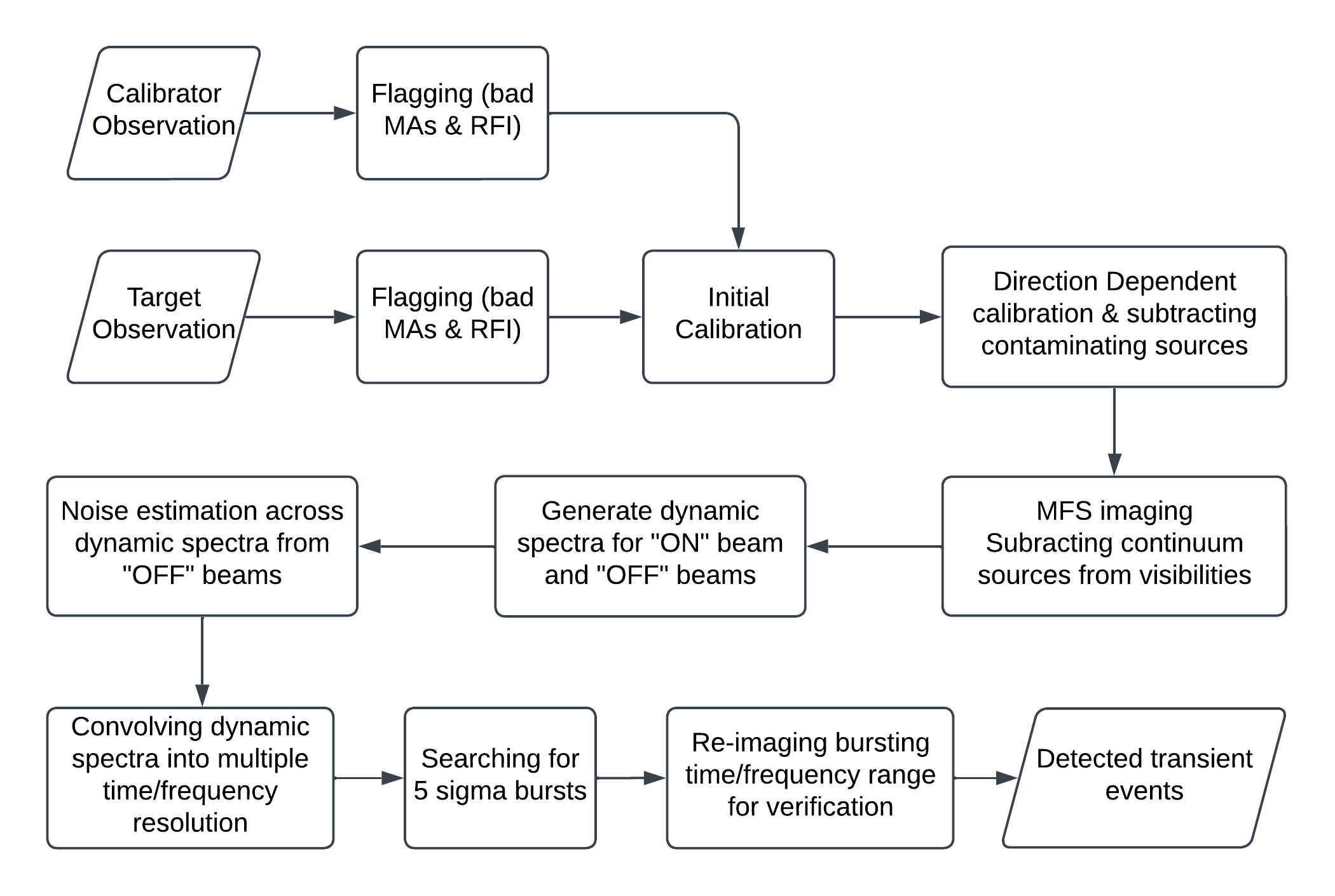}
\caption{Flowchart illustrating the stages of the data processing pipeline employed in the HD~189733 observational campaign with NenuFAR.}
\label{fig:pipe}
\end{figure*}

Our data processing pipeline\footnote{\url{https://github.com/zhangxiang-planet/exo_img_pipe}} integrates a suite of software tools designed for NenuFAR and LOFAR datasets, leveraging the capabilities of Nenupy \citep{alan_loh_2020_4279405}, Nenucal\footnote{\url{https://gitlab.com/flomertens/nenucal-cd}} \citep{2024A&A...681A..62M}, DP3 \citep{2018ascl.soft04003V}, AOFlagger \citep{2012A&A...539A..95O}, DDFacet \citep{2018A&A...611A..87T}, KillMS \citep{2014A&A...566A.127T,2015MNRAS.449.2668S, 2023ascl.soft05005T}, DynspecMS\citep{Tasse2025}, and WSClean \citep{offringa-wsclean-2014}. Prior to processing with our pipeline, initial preprocessing steps were performed following standard NenuFAR routines, including averaging and initial flagging of radio frequency interference (RFI). Subsequently, our pipeline executes the following comprehensive sequence: (1) Identification and removal of "bad" MAs at time of observation (due to temporary issues such as electronic failure or maintenance) and of RFI; (2) Initial calibration using observations of known calibrators; (3) Direction-Dependent (DD) calibration for mitigating pollution by prominent A-team sources (Cyg A, Cas A, Tau A and Vir A \citep{2020A&A...635A.150D}); (4) Dynamic spectra generation across various directions within the FoV; (5) Application of source-finding algorithms in the dynamic spectra to identify relevant signals. A step-by-step depiction of these pipeline stages is presented in Figure \ref{fig:pipe}.

\subsection{The Pipeline}

First, flagging operations are applied to both the calibrator and the target data to ensure quality. Utilizing a predefined list of channels known to be contaminated by RFI from previous observations, we proactively flag these channels. Additionally, AOFlagger is employed for automatic RFI detection and flagging. The identification of malfunctioning MAs involves an iterative approach: an extra calibration round using the calibrator data precedes the examination of the calibration solutions, considering both amplitude and phase. Should the calibration solutions reveal significant discontinuities for any MA, the MA is designated as faulty, leading to the flagging of all associated baselines. On average, approximately 7\% of MAs were flagged during the observations.

Following the flagging process, we proceed with the initial full-Stokes calibration, which is direction-independent and utilizes observations of calibrators. This calibration is done with DP3, leveraging models of A-team sources comprising of point sources and Gaussians. The amplitude/phase calibration solutions derived from this step are then transferred to and applied on the target observations.

Then, we apply full-Stokes direction-dependent calibration to mitigate the influence of contaminating A-team sources on the visibilities. Given their prominence in the low-frequency radio sky, A-team sources can degrade data quality even when located within the sidelobes outside the FoV. Direction-dependent calibration is beneficial at low frequencies due to the pronounced spatial variation in calibration solutions, which arises from ionospheric disturbances and the receiver beam's shape \citep{2011A&A...527A.107S}. The direction-dependent calibration process unfolds in stages: initially, DDFacet is utilized to produce a deconvoluted ("cleaned") image of the FoV, from which a sky model of sources within the FoV is derived. Subsequently, models of the A-team sources are incorporated into this sky model. Calibration tailored to multiple directions is then conducted using KillMS, enabling the subtraction of A-team source contributions from the visibilities, and isolating sources of moderate flux for further analysis.

To facilitate the search for transients within our dataset, we first generate the residual visibilities by removing the detected radio continuum sources. This involves conducting Multi-Frequency Synthesis (MFS) imaging across the entire bandwidth, to create a comprehensive sky model of the continuum sources within our FoV via clean deconvolution. Following this, we subtract these continuum sources from the visibilities, thus obtaining the residual visibilities primed for the analysis and detection of transients in the time-frequency plane.

% We created Stokes IQUV dynamic spectra for approximately 500 directions within the FoV using DynspecMS, capitalizing on NenuFAR's expansive FoV. Our analysis extended beyond the primary target to include other scientifically significant sources, such as additional exoplanets and ultracool dwarfs (UCDs). The selection of exoplanets was informed by the NASA Exoplanet Archive, whereas the UCDs were identified through the GAIA Data Release 3 (DR3) \citep{2016A&A...595A...1G, 2023A&A...674A...1G}. To comprehensively monitor for sporadic transient events, we also produced dynamic spectra for a set of field grid directions, systematically spaced one degree apart, employing the Fibonacci lattice method for even distribution \citep{2006QJRMS.132.1769S}. Although dynamic spectra were created for all the Stokes parameters, our analysis specifically focused on the Stokes I and V components. This selection is predicated on the expectation that CMI emissions, which we aim to detect, are highly circularly polarized, making the Stokes V spectra particularly relevant for our studies. A graphical representation mapping all dynamic spectra directions is presented in Figure \ref{fig:map}.

\begin{figure}
\centering
\includegraphics[width=\hsize]{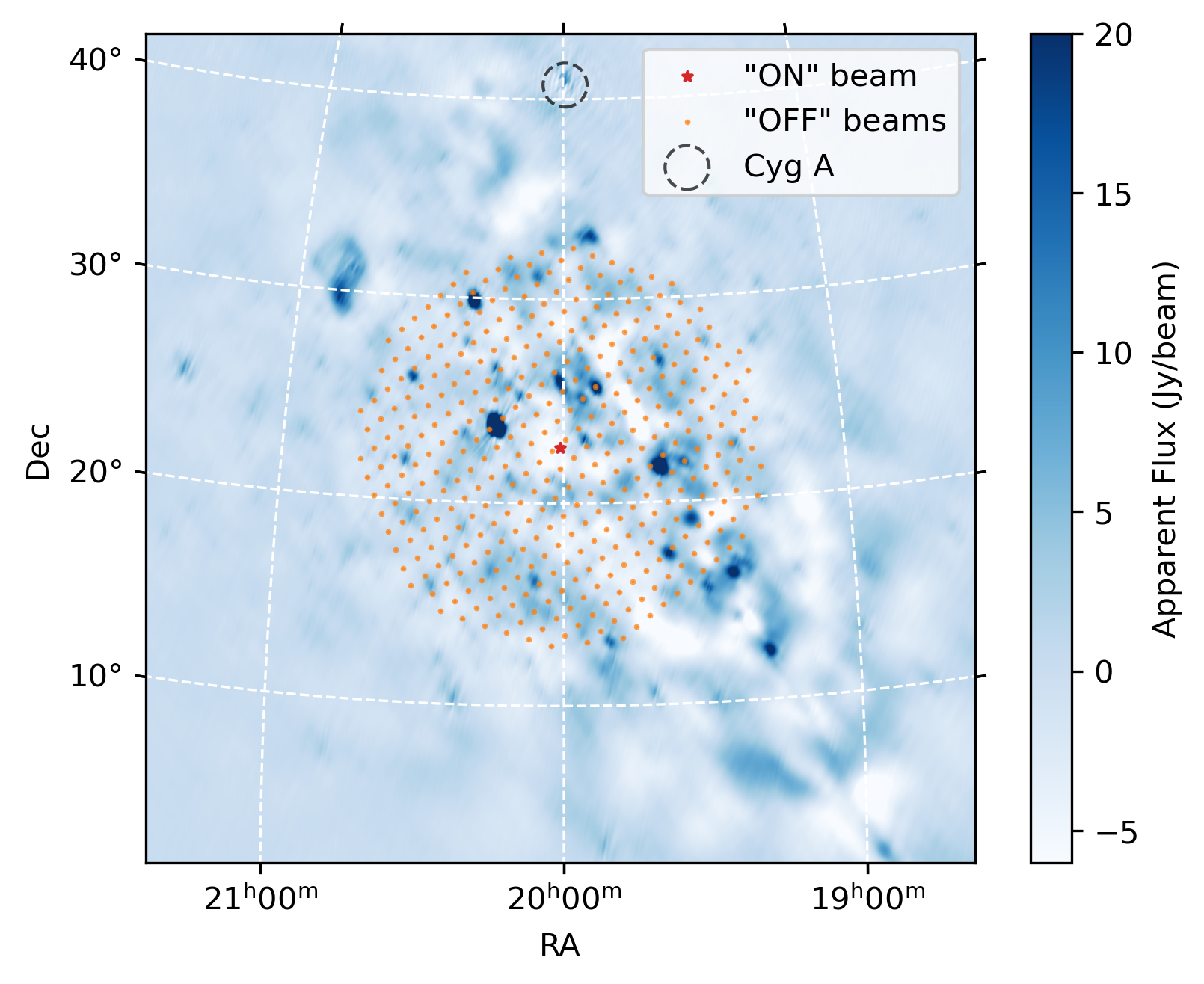}
\caption{Spatial distribution of dynamic spectra directions within the FoV, illustrated against a blue background created from a full-band MFS image prior to continuum source subtraction. The central red dot denotes the "ON" beam, located in the direction of HD~189733. Orange dots represent the "OFF" beams, whose dynamic spectra are used for noise estimation. The black circle at the top marks the location of Cygnus A, which has been subtracted from visibilities.}
\label{fig:map}
\end{figure}

We generated Stokes IQUV dynamic spectra from residual visibilities with DynspecMS, for the target direction (referred to as the "ON" beam) and for approximately 400 off-target directions (referred to as "OFF" beams) within NenuFAR’s FoV. The "OFF" beams, which are used to estimate the noise level\footnote{We also examined the OFF beams as part of a broader transient search, though such events are outside the scope of this paper and will be discussed in future work.}, were systematically spaced one degree apart using the Fibonacci lattice method for even distribution \citep{2006QJRMS.132.1769S}. Although dynamic spectra were created for all the Stokes parameters, our analysis specifically focused on the Stokes I and V components. This selection is based on the fact that CMI emissions are highly circularly polarized, making the Stokes V spectra particularly relevant for our study. Additionally, the noise level in Stokes V spectra is significantly less compared to Stokes I, further improving detection sensitivity. Nevertheless, Stokes Q and U were also examined for verification purposes in the event of potential detections. A graphical representation mapping all beam directions is presented in Figure \ref{fig:map}.

% \begin{figure}
% \centering
% \includegraphics[width=\hsize]{figures/target_distri.png}
% \caption{Spatial distribution of dynamic spectra directions within the FoV, illustrated against a blue background created from a full-band MFS image prior to continuum source subtraction. The central red dot denotes the primary target, HD~189733. Surrounding orange dots mark the locations of additional exoplanets, and yellow dots indicate UCDs. Grey dots represent evenly spaced field grids, positioned one degree apart, designed for transient event detection. The black circle at the top marks the location of Cygnus A, which has been subtracted from visibilities.}
% \label{fig:map}
% \end{figure}

The target dynamic spectra derived directly from the visibilities require a ‘flattening’ process to standardize the noise levels, which fluctuate across time and frequency. To achieve this, we first computed the mean and standard deviation (STD)\footnote{We also tested normalization using the median and the median absolute deviation (MAD), but ultimately chose the mean and standard deviation. While MAD-based methods are more robust to outliers, they complicate the interpretation of signal-to-noise ratios, especially in non-Gaussian noise conditions.} for each time/frequency cell across approximately 400 "OFF" beams in the dynamic spectra, leveraging the consistency of apparent flux noise levels across various directions. Dynamic spectra corresponding to the scientific target were excluded from these calculations to avoid including genuine signals in our noise estimation. Finally, we normalized the noise levels by subtracting the mean noise level at each time-frequency pixel (averaged across all OFF beams) and dividing by the corresponding standard deviation.

Given the uncertain time/frequency widths of potential bursts, we applied a convolution process to the normalized dynamic spectra using Gaussian kernels across a range of resolutions. Time resolutions were set on a logarithmic scale, ranging from 16 seconds to 4096 seconds with a factor of 2, while frequency resolutions also followed a similar logarithmic progression, from 240~kHz to 15.36~MHz. For each combination of time and frequency resolution, we recalculated the mean and STD dynamic spectra across the "OFF" beams. Subsequently, we computed Signal-to-Noise Ratio (SNR) dynamic spectra by subtracting the mean from the convolved dynamic spectra and dividing with the corresponding STD. 

% Our search criteria for individual transient candidates within these SNR dynamic spectra involved identifying 5 sigma events based on two key conditions: (1) a candidate's SNR must exceed 5, ensuring its impact is localized to the target direction — this helps distinguish genuine events from RFI bursts or sidelobe contamination from bright sources, which would influence a broader swath of the sky; and (2) the candidate must surpass the 5 sigma threshold within its respective dynamic spectrum, confirming its nature as a time/frequency transient rather than a bright continuum source.

Our search for individual transient candidates within the SNR dynamic spectra focused on identifying events with an SNR exceeding 5. This threshold ensures that the detected event is localized to the target direction, helping to distinguish genuine transients from RFI bursts which would affect a broader area of the sky.

Upon identifying a transient candidate, we conduct a verification process by re-imaging the specific time/frequency range of the candidate using WSClean\footnote{As indicated by the grey boxes in Figure \ref{fig:dynamic_spec}.}, generating an all-sky image including sidelobe area. This step helps rule out RFI contamination, ionospheric scintillation, sidelobe contamination, or noise peaks. Our approach follows a method similar to that described by \citet{Tasse2025}, who highlight the potential biases introduced by time-frequency weighting and demonstrate a framework to assess detection reliability using re-weighted PSF testing. To ensure thoroughness, we employ a clustering algorithm that expands to include neighboring time/frequency cells exhibiting SNRs greater than 3, adjacent to the initial bursting range. Bright RFI is typically identified and verified within the dynamic spectra, appearing as either narrow-banded signals at specific frequencies or as broadband signals with very short timescales and high brightness. Fainter RFI signals can be identified through re-imaging, as RFI typically affects multiple positions or even the entire FoV. Ionospheric scintillations are identified by re-imaging and checking for evidence of strong scintillation across the sky region, as they generally affect multiple sources rather than a single source. Sidelobe contamination can be identified by checking whether known bright sources appear at positions corresponding to sidelobes. Noise peaks are identified by re-imaging the suspected event; if no source resembling the PSF shape is visible in the image at the location of our target, the event is classified as an artifact. In cases where the imaged source matches the PSF, the event is confirmed as a transient candidate.

\subsection{Data quality}
\label{sec:quality}

Before performing scientific analysis, we assessed the quality of the data. Out of the 103 hours of observations collected during our campaign, 16 hours were deemed to be of poor quality, primarily due to thunderstorms and temporary instrumental failures. As a result, only 87 hours of data were retained for further analysis. Additionally, RFI flagging revealed that channels below 21~MHz were heavily contaminated by RFI, leading to their exclusion from further processing.

Given our focus on detecting potential CMI emissions, which are expected to exhibit strong circular polarization, it was essential to evaluate the level of polarization leakage in our data. Leakage occurs when signals from one polarization state contaminate another, potentially causing unpolarized sources to appear polarized or distorting the measured polarization fraction of polarized sources. In circular polarization, leakage typically arises in two ways: from total intensity into circular polarization, or between linear and circular polarization \citep[often referred to as the "XY phase" issue]{1996A&AS..117..149S, 2017PASA...34...40L}. To quantify the leakage from total intensity into Stokes V, we compared pixel values of bright 3C sources in Stokes V and Stokes I images, estimating a leakage factor of approximately 0.75\%. For leakage between linear and circular polarization, we measured the XY phase using observational data collected during a Jupiter burst, which is a strong polarized source below 40~MHz. At an SNR of about 1800, no significant leakage between linear and circular polarization was detected.

\begin{figure}
\centering
\includegraphics[width=\hsize]{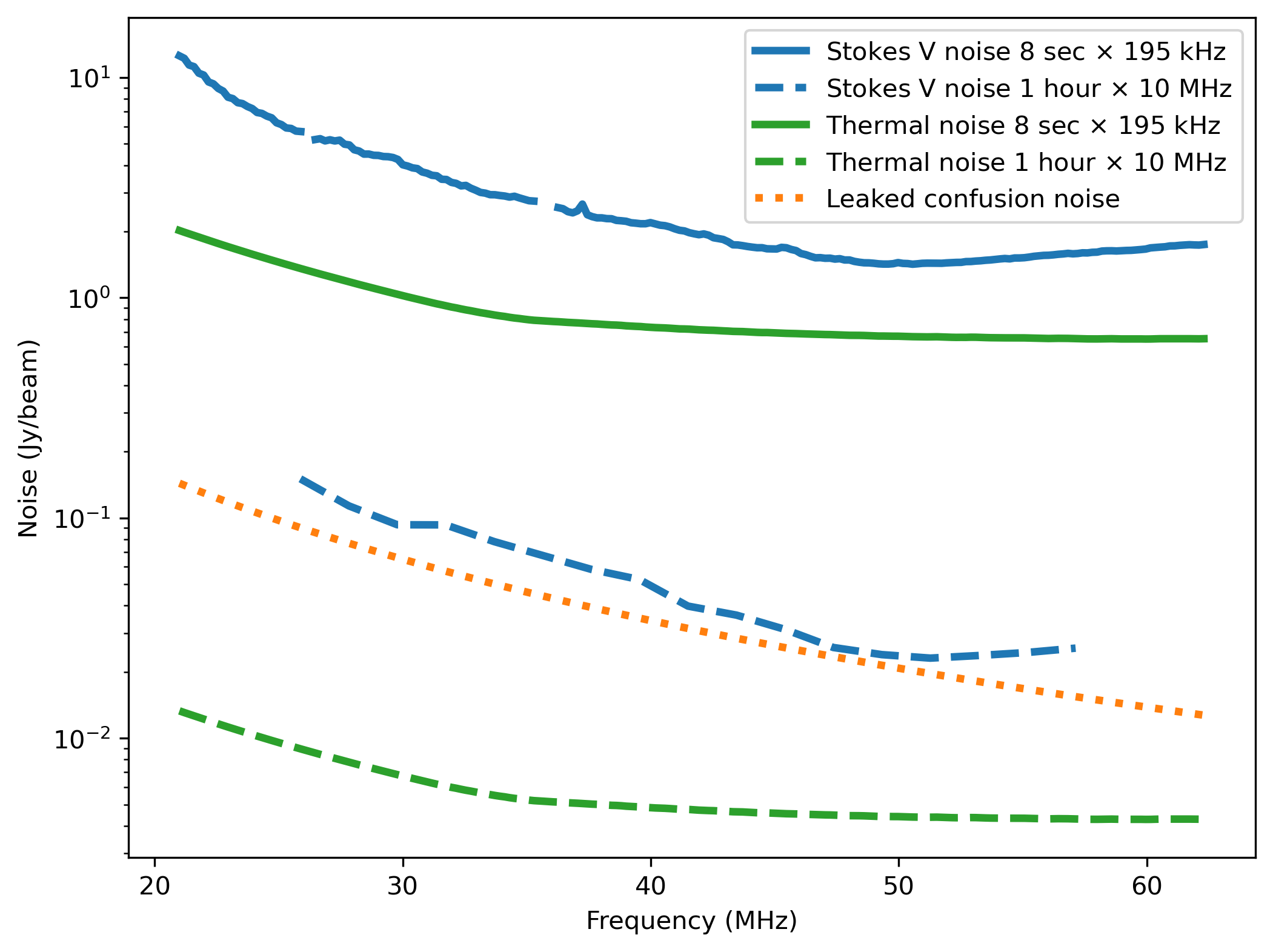}
\caption{Noise levels in the Stokes V data as achieved by the pipeline. The blue lines represent the measured noise levels from the observational data for two time-frequency integrations (8 sec $\times$ 195~kHz and 1 hour $\times$ 10~MHz), derived from both dynamic spectra and images. The green lines indicate the theoretical thermal noise of NenuFAR for the same time-frequency integrations. The orange line represents the estimated confusion noise in Stokes V, caused by the 0.75\% leakage from Stokes I.}
\label{fig:noise}
\end{figure}

Following the leakage estimation, we assessed the noise level in our Stokes V data using both images and dynamic spectra, and compared these results with the expected thermal noise. At high time-frequency resolutions (e.g., 8 sec $\times$ 195~kHz), the noise in the Stokes V data is primarily dominated by thermal noise. The lowest noise level (1.42 Jy/beam), about twice the theoretical thermal noise (0.67 Jy/beam), was observed around 50MHz. At 21MHz, the noise increased to nearly six times the thermal noise, likely due to ionospheric disturbances that are more pronounced at lower frequencies. However, we also note that leakage from Stokes I into Stokes V introduces additional confusion noise. For long integrations over wide frequency bands (e.g., 1 hour $\times$ 10~MHz), the noise level closely matches the estimated leaked confusion noise, making it challenging to detect signals below 50 mJy in our observations. Figure \ref{fig:noise} illustrates the estimated noise levels for our observations. The leaked confusion noise was calculated by multiplying the leakage factor by the theoretical confusion noise in total intensity.

Besides the analysis presented in this work, the reliability of our data processing pipeline and transient search methodology is further supported by detections made in other projects using NenuFAR. Using the same pipeline, we successfully identified broadband, polarized emissions from Starlink satellites, demonstrating their impact on low-frequency radio astronomy \citep{2024A&A...689L..10B,2025arXiv250410032Z}. Additionally, we have detected low-frequency radio bursts from other stellar and substellar systems, including repeating bursts likely originating from brown dwarfs (Zhang et al., in prep.), reinforcing the robustness of our approach when applied to astrophysical targets.

\section{Results}
\label{sec:results}

Over the course of 87 hours of observations, only one transient candidate successfully passed the verification process described earlier\footnote{A total of 16 transient candidates were initially identified, of which 15 were rejected due to likely RFI or calibration artifacts.}. This event is a 1.5 Jy circularly polarized burst, detected in the Stokes V dynamic spectrum of HD~189733 and confirmed through imaging. In addition to the search for individual bursts, we conducted a statistical analysis of possibly weaker emission from the system using the Lomb-Scargle method, aimed at identifying potential periodic signals. However, no significant periodicity was detected.

\subsection{Individual Burst}

\begin{figure*}
\centering
\includegraphics[width=\textwidth]{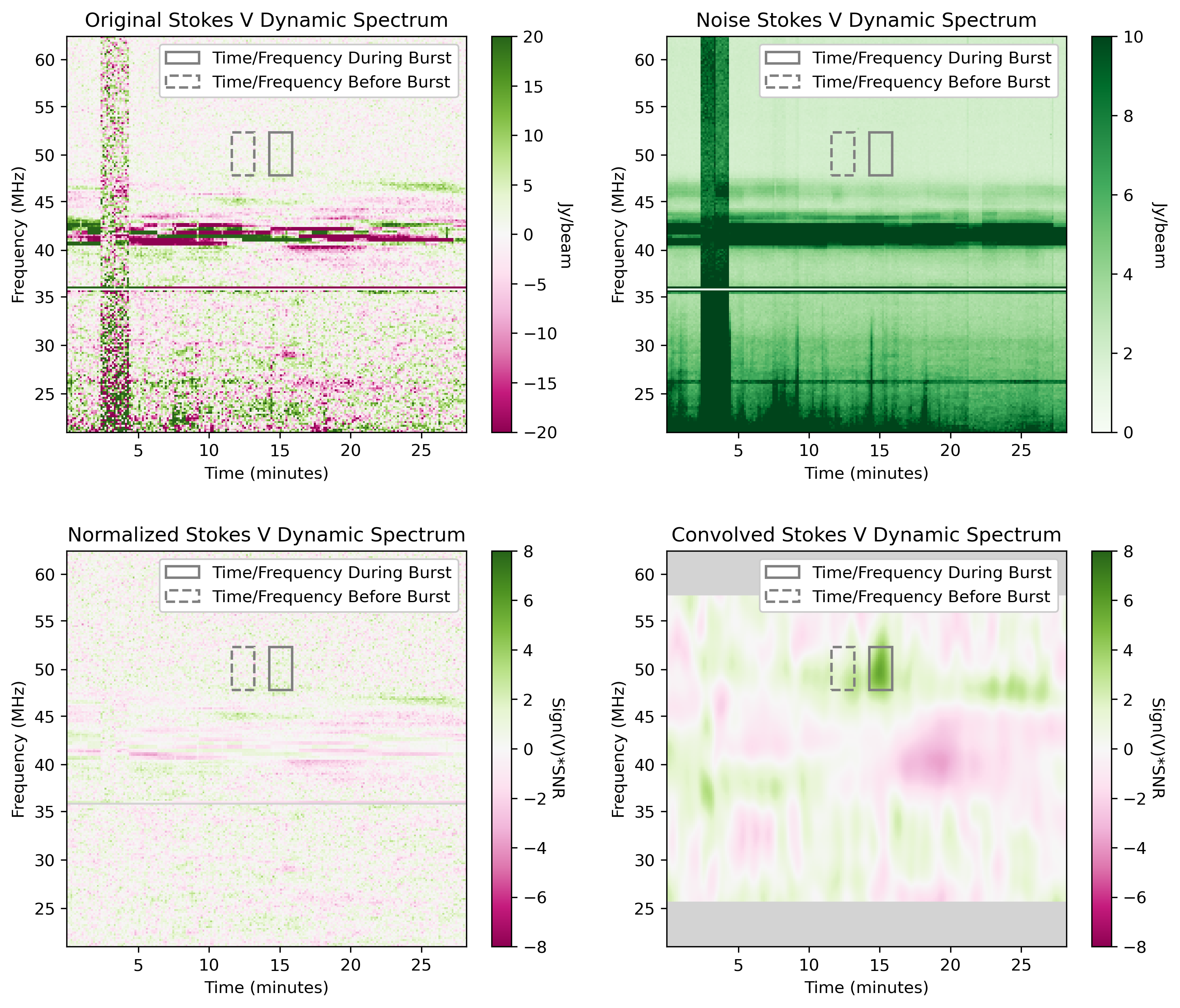}
\caption{Dynamic spectra of the HD~189733 field, showing the detection of a burst after convolution. \textbf{Top-left panel}: Original Stokes V dynamic spectrum of HD~189733, at a resolution of 8 sec $\times$ 195~kHz. The frequency range 41–43~MHz and the time range 2–5 min are affected by RFI or temporary instrumental issues. The solid grey box marks the time-frequency region of the detected burst, though it is not visible at the native resolution. A dashed grey box, located at the same frequencies but earlier in time, serves as a control region for comparison. \textbf{Top-right panel}: Noise dynamic spectrum, showing the standard deviation computed from $\sim$400 off-beam directions for each time-frequency pixel. RFI-contaminated regions appear with elevated noise. \textbf{Bottom-left panel}: Normalized Stokes V dynamic spectrum after applying noise standardization. This process flattens the noise level across the spectrum and down-weights contaminated regions. \textbf{Bottom-right panel}: Convolved Stokes V dynamic spectrum with a resolution of 40 sec $\times$ 4~MHz. The burst becomes clearly visible in the solid grey box, while no signal is seen in the control region (dashed grey box). Edge frequencies are excluded due to insufficient convolutional data.}
\label{fig:dynamic_spec}
\end{figure*}

The burst was detected in the Stokes V dynamic spectra on 2023-09-28 at 21:18 UTC, when the exoplanet was at an orbital phase of 0.0197 (close to transit). At the original dynamic spectrum resolution of 8 sec $\times$ 195~kHz, the burst was not detectable. However, after convolving the dynamic spectra to a lower time-frequency resolution of 40 sec $\times$ 4~MHz, the burst exceeded the 5 $\sigma$ threshold (Figure \ref{fig:dynamic_spec}). The emission was detectable for approximately 96 seconds and spanned the frequency range 47.6–52.1~MHz. The actual time-frequency extent may be larger, as only the peak of the burst may have been detected due to sensitivity limits. 

\begin{figure*}
\centering
\includegraphics[width=\textwidth]{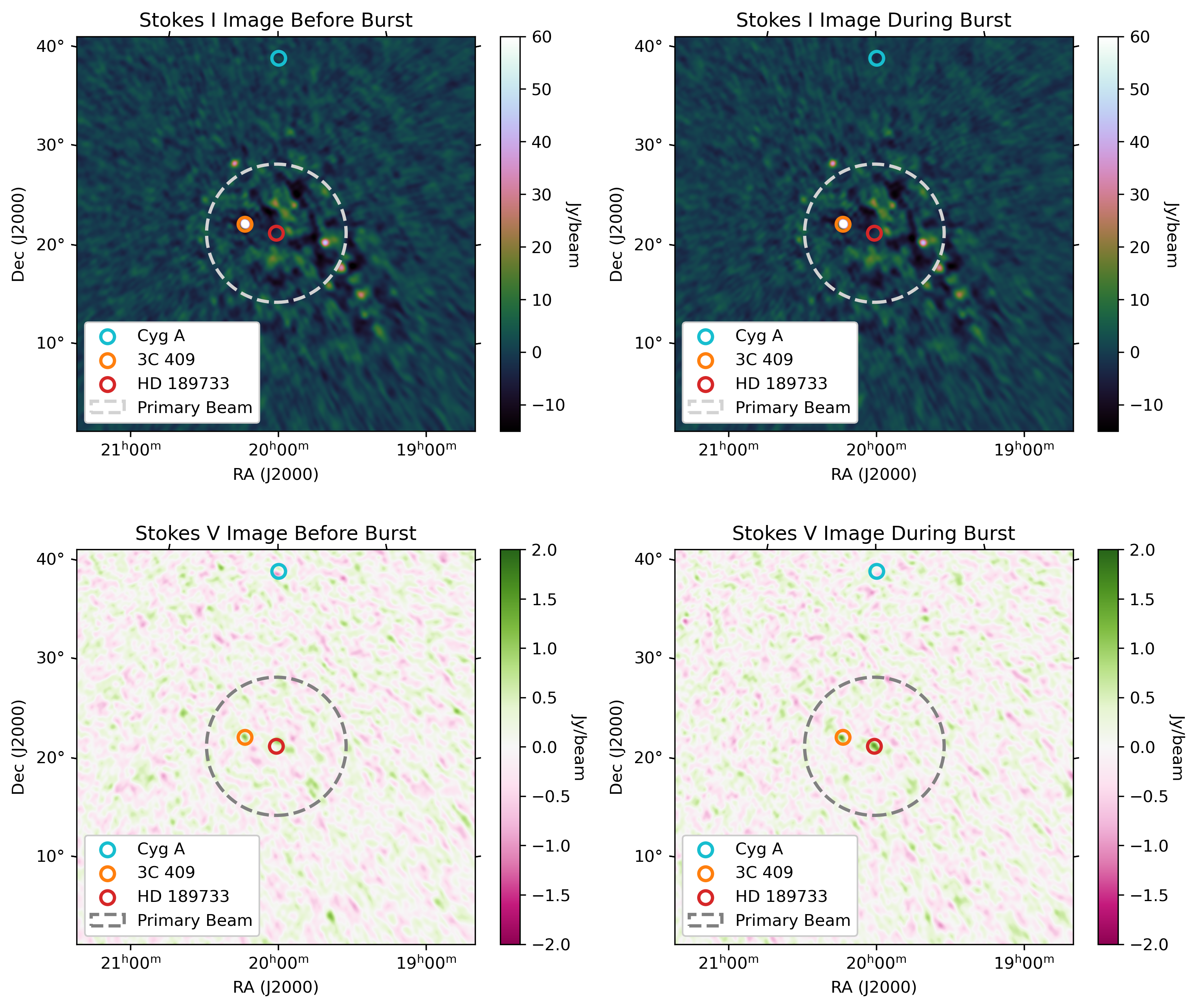}
\caption{Imaging verification of a 6$\sigma$ burst in the direction of HD~189733 with NenuFAR. All panels show apparent flux images to ensure consistent noise levels beyond the primary beam. \textbf{Top-left panel}: Stokes I image of the FoV before the burst, generated from visibilities corresponding to the control time-frequency range (dashed grey boxes) in dynamic spectra. The orange circle marks the 200 Jy radio source 3C 409, while the red circle marks the location of HD~189733. The cyan circle marks the location of Cyg A, which was subtracted from the visibilities. The large dashed grey circle represents the primary beam area of NenuFAR at 50~MHz (the central frequency of the detection). \textbf{Top-right panel}: Stokes I image corresponding to the burst's time-frequency range (as indicated by the solid grey boxes in the dynamic spectra). Due to the relatively high noise level in Stokes I, no significant emission is detected at the location of HD~189733, while the bright source 3C 409 is clearly visible. \textbf{Bottom-left panel}: Stokes V image of the FoV generated from the control time-frequency range. The orange circle indicates the leakage at the location of 3C 409, while no significant emission is detected at the position of HD~189733 (red circle). \textbf{Bottom-right panel}: Stokes V image corresponding to bursting time-frequency range. The leakage at the location of 3C 409 is still visible (orange circle), and the red circle highlights a burst detected at the location of HD~189733.}
\label{fig:reimage}
\end{figure*}

Re-imaging the data over the time and frequency range of the emission revealed a 6 $\sigma$ circularly polarized burst with a flux density of 1.5 Jy at the location of HD~189733 (Figure \ref{fig:reimage}), confirming the detection. No corresponding burst was detected in the corresponding Stokes I image, indicating that the burst is highly circularly polarized. The noise level in the Stokes I image is approximately 1.3 Jy/beam, versus 0.25 Jy/beam in Stokes V\footnote{The noise level is measured in images with a Briggs weighting of 0.5.}. Assuming that no detection means the Stokes I emission from HD~189733 is below the 3 $\sigma$ threshold, the 1.5 Jy burst in Stokes V corresponds to a minimum fractional circular polarization of 38\%. Additionally, leaked signals from bright Stokes I sources (such as 3C 409) were observed. Assuming that the bright background sources are unpolarized, we estimate the leakage level to be approximately 0.73\%, ruling out instrumental leakage as the origin of the burst.

To investigate possible multi-wavelength counterparts, we searched the TESS archive\footnote{\url{https://archive.stsci.edu/missions-and-data/tess}} for optical observations but found no available data at the time of detection. A search in the HEASARC archive\footnote{\url{https://heasarc.gsfc.nasa.gov/docs/archive.html}} similarly yielded no overlapping UV/X-ray observations.
We also checked the CHEOPS Science Archive\footnote{\url{https://cheops-archive.astro.unige.ch/archive_browser/}}, which provides high-precision photometry of exoplanet host stars, but found no observations of HD~189733 around the burst time. Additionally, no relevant light curves were available from the NASA Exoplanet Watch\footnote{\url{https://exoplanets.nasa.gov/exoplanet-watch/}}, a citizen science project that collects optical transit data from a global network of amateur and professional astronomers.

\subsection{Search for Periodicity}

The detectability of cyclotron maser emissions depends strongly on the viewing geometry, as the emission is highly beamed along a thin hollow cone \citep{2017A&A...604A..17M,2023pre9.conf03091L}. Detecting periodicity in the dynamic spectra could provide insights into the burst’s origin, particularly if the detected periods match known system timescales, such as the orbital period of the planet or the synodic period between it and the rotation of the stellar magnetic field \citep{2025arXiv250318733L}. To investigate this, we applied the Lomb-Scargle periodogram technique \citep{2018ApJS..236...16V} to all convolved dynamic spectra across multiple time-frequency resolutions. False alarm probabilities (FAPs) were estimated using the Baluev method \citep{2008MNRAS.385.1279B}, which utilizes extreme value theory to provide an analytic upper limit for the FAP, reducing the need for extensive Monte Carlo simulations.

We searched for periodic signals over the range of 0.74–35.82 days, covering key system timescales:
\begin{itemize}
    \item The minimum period (\textbf{0.74 days}) corresponds to one-third of the planet’s orbital period.
    \item The maximum period (\textbf{35.82 days}) is three times the stellar equatorial rotation period.
    \item This range includes important system periods:
    \begin{itemize}
        \item Planetary orbital period: \textbf{2.22 days}
        \item Stellar equatorial rotation period: \textbf{11.94 days}
        \item Synodic period between the planet and star: \textbf{2.73 days}
        \[
        P_{\rm synodic} = \frac{1}{\left| \frac{1}{P_{\rm planet}} - \frac{1}{P_{\rm star}} \right|}
        \]
        \item Harmonic period between the planet and star: \textbf{1.87 days}
        \[
        P_{\rm harmo} = \frac{1}{\left( \frac{1}{P_{\rm planet}} + \frac{1}{P_{\rm star}} \right)}
        \]
    \end{itemize}
\end{itemize}
Additionally, since the host star undergoes differential rotation, this range includes its polar rotation period (16.53 days) \citep{2010MNRAS.406..409F} as well.

\begin{figure*}
\centering
\includegraphics[width=\textwidth]{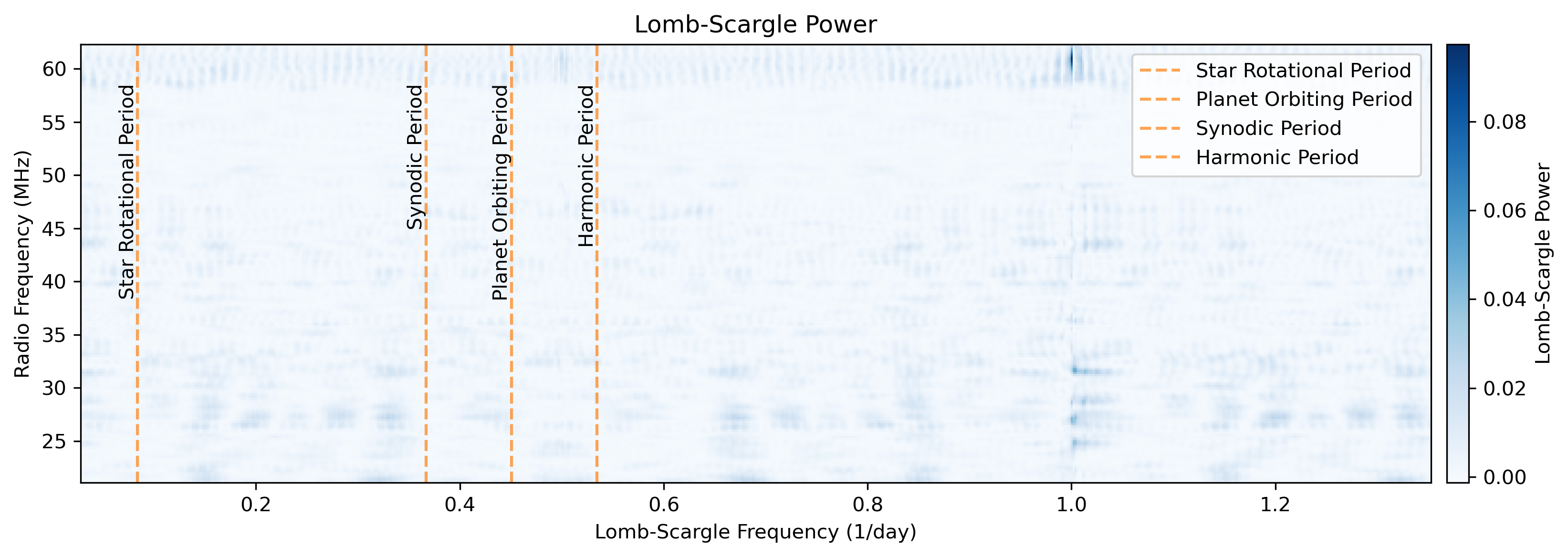}
\caption{Lomb-Scargle power distribution as a function of period (represented by the Lomb-Scargle frequency on the x-axis) and radio frequency (y-axis), based on the Stokes V dynamic spectra at the highest time-frequency resolution (8 sec $\times$ 195~kHz). Orange dashed lines mark periods of scientific interest: the stellar rotation period, the orbital period of the planet, and the synodic period and the harmonic period between the two. The peaks near 1 cycle/day are likely caused by observational windowing. Only peaks with a low false alarm probability (FAP $<$ 0.01) are considered significant.}
\label{fig:period}
\end{figure*}

No periodic signals corresponding to the planetary orbit, stellar rotation, or expected star-planet interaction timescales were detected. The only significant periodicity appeared at 60~MHz with a period of 24 hours, probably caused by RFI and/or observational windowing, rather than an astrophysical source. The Lomb-Scargle power distribution and corresponding FAP analysis are shown in Figure \ref{fig:period}.

\section{Discussion}
\label{sec:discussion}

Identifying the origin of a single 6$\sigma$ burst without an accompanying periodic signal is a challenge. To assess possible explanations, we examine the burst in the context of different emission mechanisms and compare theoretical predictions with the observed properties. We first explore whether the burst could be attributed to star-planet interaction, followed by an analysis of intrinsic stellar radio activity. We then investigate the possibility that the burst originated from a background source along the line of sight, before considering the potential for a false positive due to noise fluctuations.

\subsection{Star-Planet Interaction}

Star-planet interactions (SPI) can produce radio bursts through two primary mechanisms \citep{2024NatAs...8.1359C,Zarka2025}. The first, sub-Alfvénic interaction, is analogous to the Jupiter-Io system. In this scenario, a planet—either intrinsically magnetized or conductive with an induced magnetosphere—orbits within the Alfvén surface of its host star, being an obstacle to the stellar plasma flow. This interaction generates Alfvén waves that propagate back to the star, accelerating electrons along magnetic field lines. Radio emission is then produced via the CMI mechanism along the stellar field lines, provided the plasma-to-cyclotron frequency ratio remains sufficiently low. The second mechanism, wind-magnetosphere interaction, is similar to Earth's or Saturn's auroral processes. Here, an intrinsically magnetized planet interacts with the stellar wind, causing magnetic reconnection mainly on the planet's one side. The accelerated electrons then travel along magnetic field lines toward the planet's magnetic poles, triggering CMI emission along the magnetic field lines connected to the poles.

Several studies have modeled SPI-induced radio emission for HD~189733~b \citep{2019MNRAS.485.4529K, 2023pre9.conf03092M, mauduit:tel-04821784}, but the predicted flux densities and frequencies vary due to uncertainties in the planet’s magnetic field strength. Assuming a planetary surface field of 10 G, \citet{2019MNRAS.485.4529K} modeled the wind-magnetosphere interaction and predicted a maximum flux of 0.1 Jy at ~25~MHz. Their results also suggest that planetary emission could only escape the system when the planet is near primary transit. 

\citet{mauduit:tel-04821784} used the planetary magnetic model from \citet{2010A&A...522A..13R} to simulate both SPI mechanisms. Their sub-Alfvénic interaction model predicts a flux of 2.6 Jy, with a maximum frequency of 65.7~MHz—compatible with our detection. In contrast, their wind-magnetosphere interaction model predicts a lower flux of 0.23 Jy at a maximum frequency of 75.5~MHz. 

Additionally, using the results from the extensive MHD simulations by \citet{2022MNRAS.512.4556S}, which predict a dissipated SPI power of $P_d \sim 10^{21}$ W, we can estimate the expected radio flux density using the radio-magnetic scaling law with the most conservative efficiency factor $\beta=10^{-4}$ \citep{Zarka2025}:

\begin{equation}
    S = \frac{\beta P_d}{\Omega d^2 \Delta f} \approx 3.6 \text{ Jy},
\end{equation}

where $\Omega=0.16$ sr is the beaming solid angle \citep{2004JGRA..109.9S15Z}, $d=19$ pc is the distance to the system, and $\Delta f \approx 50$~MHz is the assumed emission bandwidth. 

Given our detected flux of 1.5 Jy, an SPI origin remains plausible, particularly via the sub-Alfvénic interaction, which predicts flux densities compatible with our observation. However, an important caveat arises from plasma conditions near the emission region. Assuming a dipolar magnetic field with a surface strength of 40 G \citep{2022MNRAS.512.4556S}, a 50 MHz emission corresponds to a height of approximately $0.3,R_\star$ above the stellar surface. At such low altitudes, the plasma density is expected to be high, resulting in a plasma-to-cyclotron frequency ratio ($f_{\mathrm{pe}}/f_{\mathrm{ce}}$) significantly greater than 0.1 — a condition that suppresses the CMI mechanism. One possible resolution is that the emission could originate from a localized region with enhanced magnetic field strength, such as a stellar magnetic hot spot. This would allow 50 MHz emission to occur at higher altitudes, where the plasma density is lower and CMI conditions are more favorable.

\subsection{Intrinsic Stellar Activity}

In addition to star-planet interactions, stars can produce intrinsic radio emissions, which are broadly classified into two categories \citep{1985ARA&A..23..169D, 1998ARA&A..36..131B, 2002ARA&A..40..217G, 2020A&A...639L...7V}:
1.	Coherent emission, including plasma emission and CMI, which typically occurs at decimeter to decameter wavelengths.
2.	Incoherent emission, such as gyrosynchrotron, free-free, and gyromagnetic processes \citep{2020FrASS...7...57N}, which primarily dominates at millimeter and centimeter wavelengths.

Since our detection was made at meter-decameter wavelengths, and gyrosynchrotron emission is generally broadband and weakly polarized \citep{2024NatAs...8.1359C}, it is unlikely to be the cause of the observed burst. Free-free emission is also an improbable explanation, as it is intrinsically weakly polarized and generally optically thick at such low frequencies. Additionally, \citet{2019MNRAS.485.4529K} predicted that the free-free emission from the stellar wind of HD~189733~A should be $\le5$ $\mu$Jy, several orders of magnitude below the 1.5 Jy burst we detected. Finally, gyromagnetic emission is not expected at the low frequencies. Given these constraints, we focus on plasma emission and CMI as the most plausible explanations.

Plasma emission, analogous to type II and type III solar radio bursts, occurs when an electron beam—accelerated by either magnetic reconnection or a shock wave—propagates into a dense thermal plasma, exciting Langmuir waves that subsequently produce radio emission \citep{2002ASSL..279.....B, 2019SoPh..294..122A}. Compared to CMI, plasma emission is generally moderately polarized, broadband, and sporadic \citep{2021MNRAS.500.3898V, 2024NatAs...8.1359C}. However, due to the absence of an accurate estimate of fractional polarization, the lack of periodicity, and the limited time-frequency resolution of our detection, distinguishing between plasma emission and CMI remains challenging.

To evaluate whether the burst could be due to plasma emission, we estimated the expected coronal scale height of HD~189733~A, following the suggestion from \citep{2024NatAs...8.1359C}. The coronal scale height can be derived from the stellar coronal temperature using the hydrostatic equilibrium model:

\begin{equation}
   H = \frac{k_B T}{\mu m_p g}.
\end{equation}

where $H$ is the coronal scale height, $k_B$ is the Boltzmann constant, $T$ is the coronal temperature, $\mu$ is the mean molecular weight (we use $\mu \approx 0.6$ in analogy to the solar corona, where it accounts for a mixture of hydrogen and helium \citep{1992ApJ...391..380A}), $m_p$ is the proton mass, and $g$ is the stellar surface gravity. 

XMM-Newton observations indicate that HD~189733~A has an average coronal temperature of 0.4 keV, increasing to 0.9 keV during flares \citep{2022A&A...660A..75P}. Using these values, along with an estimated electron density of \( n_e = (3 - 10) \times 10^{10} \) cm\(^{-3} \) from earlier observations \citep{2014ApJ...785..145P}, we estimate the height of 50~MHz plasma emission to be \( h_{50} \approx 2.4 R_\star \) for the quiescent state and \( h_{50} \approx 5.7 R_\star \) for the flaring state, adopting a stellar radius of \(R_* \approx 0.8 R_{\odot} \approx 5.43 \times 10^5\) km \citep{2015MNRAS.447..846B}. These values suggest that, although plasma emission at 50~MHz is possible, it would require an unusually extended and dense corona or a transient energetic event, such as a powerful flare-driven shock, to sustain the necessary plasma conditions at such large heights.

We also estimated the expected height of CMI emission if it were responsible for the burst. Assuming a dipolar stellar magnetic field, the field strength at height \( h \) above the stellar surface follows:

\begin{equation}
   B(h) = B_0 \left( \frac{R_\star}{R_\star + h} \right)^3,
\end{equation}

where \( B(h) \) is the magnetic field strength at height \( h \), \( B_0 = 40 \) G\footnote{\citet{2010MNRAS.406..409F, 2017MNRAS.471.1246F} estimated \( B_0 \) for HD~189733~A from early studies. However, a recent ZDI magnetic map around the epoch of our observations suggests a lower \( B_0 \) of approximately 10–20 G (A. Strugarek, priv. comm.), though this does not affect our conclusions.} is the surface magnetic field strength, and \( R_\star \) is the stellar radius. For CMI at 50~MHz, the required emission height is \( h_{50} \approx 0.3 R_\star \), where the plasma density is expected to be high. This raises a challenge for CMI, as the emission requires a plasma-to-cyclotron frequency ratio \(f_{\mathrm{pe}} / f_{\mathrm{ce}} \lesssim 0.1\). One possible resolution is that the burst originated from a localized region with stronger-than-average magnetic fields, such as a stellar magnetic hot spot. This would shift the emission to higher altitudes, where plasma densities are lower and CMI conditions more favorable. This scenario is plausible, particularly if the burst coincided with magnetic activity such as flares or reconnection events, which are known to transiently restructure coronal magnetic topology.

To further distinguish between CMI and plasma emission, we estimated the brightness temperature of the detected burst. The 50\,MHz emission from HD\,189733 exhibits a flux density of approximately 1.5\,Jy. Using the Rayleigh-Jeans approximation, the brightness temperature \(T_B\) is given by:

\begin{equation}
    S = \frac{2 k_B T_B}{A} \left(\frac{\omega}{\Omega}\right) = \frac{2 k_B T_B \omega}{\lambda^2},
\end{equation}

where \(S\) is the observed flux density, \(A = \pi R_*^2\) is the emitting surface area, \(d = 19\) pc is the distance to the star, and \(\omega = A / d^2\) is the solid angle subtended by the source. Thus we obtain \(T_B \approx 7 \times 10^{15}\)~K.

According to \citet{2021MNRAS.500.3898V}, CMI can generate brightness temperatures up to \(10^{15}\)~K due to the efficient amplification of waves in regions with an inverted electron distribution, typically a loss-cone distribution. For comparison, CMI-generated radio bursts from Jupiter reach even higher brightness temperatures, on the order of \(10^{18}\)-\(10^{20}\)~K \citep{2007P&SS...55..598Z}. In contrast, plasma emission is constrained by the available energy in Langmuir waves and typically saturates at \(T_B \sim 10^{12}\)~K under typical coronal conditions. 

These results suggest that the burst could originate from intrinsic emission of HD\,189733\,A, with both CMI and plasma emission remain viable mechanisms for the detected burst. However, each mechanism would require specific physical conditions to operate effectively at the observed frequency and intensity, such as a localized magnetic enhancement for CMI or an extended, dense coronal structure for plasma emission. A summary of how each proposed mechanism compares against key observational and theoretical constraints is provided in Table~\ref{tab:mechanisms}.

\begin{table*}
\caption{Comparison of potential emission mechanisms with key observational and physical constraints.}
\label{tab:mechanisms}
\centering
\begin{tabular}{l c c c c}
\hline\hline
Mechanism & Power & $f_{\mathrm{ce}}$ & $f_{\mathrm{pe}}/f_{\mathrm{ce}}$ & Viewing angle \\
\hline
Sub-Alfvénic SPI & Yes & Yes & Plausible (hotspot)$^\dagger$ & Marginal$^\ddagger$ \\
Wind-magnetosphere & Plausible & Unknown & Unknown & Favorable$^\ddagger$ \\
Stellar CMI & Yes & Yes & Plausible (hotspot)$^\dagger$ & Irrelevant \\
Stellar plasma emission & Plausible (flare) & Irrelevant & Irrelevant & Irrelevant \\
\hline
\end{tabular}
\tablefoot{
\tablefoottext{$^\dagger$}{A low $f_{\mathrm{pe}}/f_{\mathrm{ce}}$ ratio is required for CMI to operate. In dense stellar coronae, this ratio is typically too high unless conditions are locally altered (e.g., magnetic hotspots).}
\tablefoottext{$^\ddagger$}{The fact that the burst was detected during the exoplanet’s transit rather than in quadrature supports the wind–magnetosphere rather than the sub-Alfvénic SPI model.}
}
\end{table*}

\subsection{Possibilities of Other Sources along the Line of Sight}

Given NenuFAR's resolution of approximately 0.4 degrees\footnote{This resolution is larger than the theoretical resolution of NenuFAR (0.1 degree) at the frequency, due to two of the four remote MAs being flagged during most of the observational campaign.} at 50~MHz, the detected burst may not necessarily originate from the HD~189733 system. To investigate this possibility, we searched the Gaia Data Release 3 (DR3) catalog \citep{2016A&A...595A...1G, 2023A&A...674A...1G} for potential sources within this angular resolution, focusing on active stars and cool stars.

For these stars, we applied two selection criteria: (1) the source is labeled as "active" in Gaia DR3, or (2) the source has an effective temperature below 6000 K, corresponding to M and K dwarfs. To exclude distant objects, we restricted the search to sources within 500 pc, resulting in 246 candidate stars. The most notable source in the range is HD~189733~B, the M dwarf companion to the primary star, while the other candidates are all at larger distances (more than 100 pc away from the Earth). 

HD~189733~B has an effective temperature of 3100 K and is located at a projected separation of 216 AU from HD~189733~A \citep{2006ApJ...641L..57B}. Given this separation, direct magnetic interaction between the two stars is negligible. However, M dwarfs are known low-frequency radio emitters \citep{2021NatAs...5.1233C}, and while most detections with LOFAR are at the mJy or sub-mJy level, some have been associated with M dwarfs at distances up to 870 pc \citep{2023A&A...678A.161G}. Since HD~189733~B is only $\sim$19 pc away, an emission at Jy levels cannot be ruled out. Future efforts including high-resolution imaging (such as with LOFAR international baselines or SKA-Low) or with simultaneous multiwavelength (e.g., X-ray) observations would be crucial to spatially and temporally disentangle the emission origin between the primary K star HD~189733~A and its M dwarf companion HD~189733~B.

For completeness, we also searched the ATNF pulsar catalog \citep{2005AJ....129.1993M} for potential pulsars within the resolution limit. No known pulsars were found in the resolution. We searched the GAIA DR3 ultracool dwarf (UCD) catalog as well, but no known UCDs were located within the resolution either.

\subsection{Possibility of a Noise Fluctuation}

A final possibility is that the detected burst is a statistical noise fluctuation rather than an astrophysical signal. Although the burst reached a significance level of 6$\sigma$, the large volume of data analyzed in this study increases the likelihood of encountering a high-significance noise peak, a phenomenon known as the "Look-Elsewhere Effect" \citep{2010EPJC...70..525G}. 

In our analysis, we considered 87 hours of data, with the time-frequency resolutions varying across convolution kernels, the highest convolved resolution being 16 sec $\times$ 240~kHz. This results in approximately 13.3 million "independent" time-frequency cells across all kernels.

We estimate the global probability, $P_{\text{global}}$, of any cell exceeding the 6$\sigma$ level as:

\begin{equation}
    P_{\text{global}} = 1 - (1 - P_{\text{local}})^N = 1 - e^{N \ln(1 - P_{\text{local}})},
\end{equation}

where $P_{\text{local}}$ is the two-tailed probability\footnote{We adopt the two-tailed probability since Stokes V signals can be either positive or negative.} of exceeding 6$\sigma$ under Gaussian noise assumptions, given by:

\begin{equation}
    P_{\text{local}} = 2 \times \left[1 - \Phi(6)\right],
\end{equation}

with $\Phi(6)$ being the Cumulative Distribution Function (CDF) of the standard normal distribution evaluated at $\sigma = 6$. Substituting the values, we find $P_{\text{local}} = 2 \times 10^{-9}$ (corresponding to a 6$\sigma$ event) and $N = 13.3$ million. Since $P_{\text{local}}$ is small, we can apply the first-order Taylor expansion:

\begin{equation}
    \ln(1 - P_{\text{local}}) \approx -P_{\text{local}}, \quad \text{for } P_{\text{local}} \ll 1,
\end{equation}

which simplifies the global probability to:

\begin{equation}
    P_{\text{global}} = 1 - e^{-P_{\text{local}} N} \approx 2.6\%.
\end{equation}

Thus, considering the large number of independent cells, the probability of a 6$\sigma$ event occurring randomly in our dataset is approximately 2.6\%. However, this number is only a rough approximation, as the estimate is further complicated by the non-Gaussian nature of the noise due to leakage effects, as well as the partial dependence between the 13.3 million cells, which overlap across different time-frequency resolutions. As a result, while the burst is formally significant, the possibility of a false positive cannot be entirely excluded.

\section{Conclusions and Perspectives}
\label{sec:conclusions}

In this work, we present the tentative detection of a low-frequency, highly circularly polarized burst from the HD~189733 system using NenuFAR. The burst, detected at 50~MHz with a flux density of 1.5 Jy and a significance of 6 $\sigma$, remains an intriguing signal. Our investigation explored multiple possible origins, including star-planet interaction, intrinsic stellar activity, background sources along the line of sight, and the possibility of a noise artifact.

The properties of the burst, particularly its high circular polarization and frequency range, are consistent with CMI emission. 
Comparisons with theoretical models suggest that multiple emission mechanisms are viable.
Sub-Alfvénic SPI and stellar CMI both match the observed frequency, but require locally low plasma density, potentially enabled by magnetic hotspots. The wind–magnetosphere scenario is energetically plausible and favored by the burst’s occurrence near transit. Plasma emission remains less likely, requiring an unusually powerful flare.
We also investigated potential contributions from background sources and found that HD~189733~B, an M dwarf companion to the primary star, could be a possible radio emitter. We also assessed the likelihood of a statistical noise peak, finding that while a false positive cannot be entirely excluded, the detection remains significant.

Future work should focus on follow-up observations to determine whether similar bursts reoccur. Detecting a second burst would help confirm the astrophysical nature of the signal, while a third would allow us to search for periodicity, providing further insights into its origin. Multi-wavelength observations, particularly in X-rays and optical, would help determine whether the star was flaring at the time of the burst. Wideband observation in radio frequencies could also help monitoring potential stellar activities. From a technical perspective, improving NenuFAR’s primary beam model—such as adopting a full embedded element model—would enhance polarization accuracy across the field of view. Additionally, collaborations with long-baseline instruments like LOFAR could refine sky modelling, improving sensitivity and reducing confusion noise.

More broadly, our study illustrates how recent technical and methodological advances may help overcome long-standing limitations in the search for exoplanetary and SPI-related radio emission. In particular, low-frequency, long-duration campaigns like this one, alongside techniques such as the DynSpecMS method, enable the detection of faint, polarized radio bursts that were previously inaccessible. Applied to large-scale surveys (e.g. LoTSS, LoTSS-Deep) and future facilities (e.g. LOFAR 2.0, SKA-Low), this approach may unlock a new population of transient emitters in the substellar regime.

More progress will require three key ingredients: (1) deeper and longer low-frequency observations with good polarization calibration; (2) improved calibration and imaging techniques to extract faint transients from wide fields; and (3) refined target selection based on Gaia catalogs, updated exoplanet databases, and flare-star monitoring. Dynamic survey instruments, such as TRON at MeerKAT, could also play a key role in real-time identification of transient sources. With the right combination of observational strategy, sensitivity, and survey depth, the coming years may finally yield the long-sought robust detections of magnetospheric and SPI-related exoplanetary radio emission.

\begin{acknowledgements}
The authors acknowledge funding from the ERC under the European Union’s Horizon 2020 research and innovation programme (grant agreement no. 101020459 - Exoradio).

This work is based on data obtained using the NenuFAR radiotelescope. NenuFAR has benefited from the funding from CNRS/INSU, Observatoire de Paris, Observatoire Radioastronomique de Nançay, Université d’Orléans, Région Centre-Val de Loire, DIM-ACAV -ACAV+ \& -Origines de la Région Ile de France, and Agence Nationale de la Recherche. We acknowledge the collective work from the NenuFAR-France collaboration for making NenuFAR operational, and the Nançay Data Center resources used for data reduction and storage.

This work has made use of data from the European Space Agency (ESA) mission
{\it Gaia} (\url{https://www.cosmos.esa.int/gaia}), processed by the {\it Gaia}
Data Processing and Analysis Consortium (DPAC,
\url{https://www.cosmos.esa.int/web/gaia/dpac/consortium}). Funding for the DPAC
has been provided by national institutions, in particular the institutions
participating in the {\it Gaia} Multilateral Agreement. 

X.Z. acknowledges the support of a postdoctoral fellowship at CSIRO, Australia, where the low-frequency polarimetry techniques developed during that time contributed to this work.

A.S. acknowledges support from the European Research Council project ExoMagnets (grant agreement no. 101125367).

J.D. acknowledges support acknowledges funding support by the TESS
Guest Investigator Program G06165 and from NASA through the NASA Hubble Fellowship grant \#HST-HF2-51495.001-A awarded by the Space Telescope Science Institute, which is operated by the Association of Universities for Research in Astronomy, Incorporated, under NASA contract NAS5-26555.

The authors acknowledge the use of AI-assisted copy editing (ChatGPT) to improve the text of the manuscript.

\end{acknowledgements}

\bibliographystyle{aa} % style aa.bst
\bibliography{reference} % your references Yourfile.bib

\end{document}